# Electrically tunable magnetism and unique intralayer charge transfer in Janus monolayer MnSSe for spintronics applications

Yu Chen,[1] Qiang Fan,[2] Yiding Liu,[3] and Gang Yao[4,*]

[1]School of Science, Inner Mongolia University of Technology, Hohhot, 010051, China

[2]School of Electronic and Material Engineering, Leshan Normal University, Leshan 614004, Sichuan, China

[3]College of Mathematics and Physics, Leshan Normal University, Leshan, 614004, China

[4]Tsung-Dao Lee Institute, Shanghai Jiao Tong University, Shanghai 200240, China

*To whom correspondence should be addressed. Email: yaogang1257@sjtu.edu.cn

**Controlling magnetism and electronic properties of two-dimensional (2D) materials by purely electrical means is crucial and highly sought for high-efficiency spintronics devices since electric field can be easily applied locally compared with magnetic field. The recently discover 2D Janus crystals has provide a new platform for nanoscale electronics and spintronics due to their broken inversion symmetry nature. The intrinsic ferromagnetic Jauns monolayer, and hence the tunable physical properties, is therefore of great interest. Here, through comprehensive density functional theory calculations and Monte Carlo simulations, we unveil that single-layer MnSSe is an intrinsic ferromagnetic half-metal with a direct band gap of 1.14 eV in spin-down channel and a Curie temperature of about 72 K. The exchange coupling can be significantly enhanced or quenched by hole and electron doping, respectively. In particular, a small amount of hole doping MnSSe can tune its magnetization easy axis in between out-of-plane and in-plane directions, which is conducive to designing 2D spin field-effect transistor for spin-dependent transport. We also find a reversible longitudinal interlayer charge transfer between S and Se layers for the first time that is highly sensitive to the applied external electric field. Interestingly, the directions of charge flow and the applied field are the same. The behavior originates from the coexistence and/or the competition of external and built-in fields. These findings, together with the excellent stability and large in-plane stiffness, can greatly facilitate the development of nanoscale electronics and spintronics devices based on 2D MnSSe crystal.**

# I. INTRODUCTION

Spintronics or spin-based electronics devices formed based on spin and charge degrees of freedom of electrons have long held promise for developing high-efficiency information technology and fundamental physics [1,2]. During the last decade, plenty of two-dimensional (2D) materials were discovered and proposed. Among all these systems, transition metal dichalcogenides (TMDs) are ones of the most typical representative because some of them host unique electronic and optical properties compared to the star material, graphene [3,4]. Although TMDs host intrinsic in-plane asymmetry, the out-of-plane mirror symmetry is considered as a disadvantage that limits the spin degrees of freedom, therefore hinders their wide applications in valleytronics, spintronics, and magneto-optoelectronics. Experimentally, by applying an external electric field [5,6] or changing the structure from a non-polar system to a poplar one, out-of-plane inversion asymmetry in 2D nanosheets was successfully achieved. For a TMDs monolayer, using a third element to replace the top or bottom layer, forming an asymmetric sandwich construction known as Janus transition-metal dichalcogenides (Janus TMDCs) monolayer [7]. The successful synthesis of MoSSe by chemical vapor deposition brought forward in recent years made the research reach a climax [8,9]. To data, the potential applications of Janus monolayer such as SnSSe, TMXY (TM= Ti, Zr, or Hf; X=S or Se; Y=O or S; X≠Y), and group-III monochalcogenide in sodium-ion batterie, ultrafast laser, photocatalysts, electronics, energy conversion, and sensor have been extensively analyzed [10-13].

With respect to spintronic applications, intrinsic ferromagnetism and its precisely and flexibly controlling are highly desirable. Recently, magnetic ordering was also predicated in Jauns monolayer TMXY (TM=Mn, V, or Cr; X/Y=S, Se, or Te; X≠Y) [14]. Then the question is that can magnetic properties be further tuned by an external method? For instance, carrier doping and electric field can efficiently induce or control the performance of 2D crystals [15-20]. Compared with strain engineering and defect engineering (*e.g.*, creating points and surface adsorption adatoms), lattice mismatch and homogeneity no longer need to be considered. In addition, electrically controllable magnetic anisotropy is highly desirable for non-volatile magnetoelectric random access memory [21]. Such an investigation in Janus monolayer, however, has rarely been reported in the literature. On the other hand, considering the sandwich-like structure with different surfaces of top and bottom, one central issue naturally arises that what the charge behaves in this non-trivial structure under external electric field. The above questions deserve a careful investigation. Also, the possible novel phenomenon will promote the development of spintronics or spin-based electronics.

In this paper, we predict in Janus monolayer MnSSe a promising electrically tunable ferromagnetism, large magnetocrystalline anisotropy, and unique intralayer charge transfer. Previous work proposed that MnSSe possesses rigid ferromagnetic order [14,22]. Our calculations suggest that pristine MnSSe is ferromagnetic half-metal with a Curie temperature of about 72 K. The magnetic coupling can be considerably enhanced or quenched with carrier doping. The magnetization can be changed from an easy plane to an easy axis with a small amount of hole doping. Further, a unique intralayer charge transfer behavior is first presented and explained based on the unusual asymmetric 2D crystal structure. The results presented are essential for electronics and spintronics applications.

## II. METHOD and COMPUTATIONAL PROCEDURES

Our calculations were performed using a plane-wave basis set and projector augmented-wave method [23,24] as implemented in the Vienna Ab initio Simulation Package (VASP) code [25]. The GGA-PBE exchange-correlation functionals [26] were used to describe the electron-ion interactions. The kinetic cutoff energy was set to 420 eV. To avoid the interlayer interactions, a vacuum spacing of 15 Å was introduced in the direction normal to the sheet. The convergence criteria for energy were set to be $1\times10^{-6}$ eV. The $3p^64s^23d^5, 3s^23p^4$, and $4s^24p^4$ states are treated as valence states for Mn, S, and Se atoms, respectively. We use the GGA+$U_{eff}$ method introduced by Dudarev *et al.*[27] to describe the localized Mn-3$d$ orbital. According to previous reports [14,28], $U_{eff}$=$U$-$J$=3 was selected, where $U$ and $J$ denote on-site Coulomb and exchange parameters and were set to 3.7 and 0.7 eV, respectively. For non-collinear calculations, the spin-orbit coupling (SOC) interaction was considered and performed on a unit cell. The Brillouin zone integration was performed using Γ-centered k-point meshes of 11×11×1 and 23×23×1 for primitive unit cell and 2×2×1 supercell, respectively. Test calculations with large energy cutoff and k-points mesh demonstrated that the results were fully converged.

Both the electron and hole doping were simulated by adding or removing electrons into the system, and a uniform background charge is adopted to maintain charge neutrality. Phonon dispersion was calculated on a 5×5×1 supercell by the finite displacement method [29] with a displacement of 0.1 Å, as embedded in the PHONOPY code [30]. To explore the magnetic coupling versus temperature Monte Carlo Metropolis algorithm simulations based on the Ising Hamiltonian model was performed. In this simulation, the spins on all magnetic sites flip randomly and a 100×100×1 supercell containing ~30000 local magnetic moments is used to reduce the periodic constraints. At each temperature, the equilibrated magnetic moment is taken after the simulations lasted for $1\times10^7$ loops. The crystal structure and charge densities were visualized by VESTA software [31]. The vaspkit program [32] was employed in both the pre- and post- data processing.

## III. RESULTS

### A. Stability of single-layer MnSSe

Janus TMDCs monolayers, which are isostructural to the TMDCs, have been known as a family of newly 2D materials since 2014 [8], in which the TM atoms sandwiched between bottom-X and top-Y layers, resulting in trigonal and octahedral prismatic coordination in 1$T$ and 2$H$ phase. For isolating single-layer MnSSe, 1$T$ structure is 400 meV lower in energy than the 2$H$ one. We henceforth only focus on the 1$T$ structure [Fig. 1(a)]. To check the chemical stability, we calculate the cohesive energy by $E_{coh}=(E_{Mn}+E_S+E_{Se}-E_{MnSSe})/3$, where $E_{MnSSe}$, $E_{Mn}$, $E_S$, and $E_{Se}$ are the energy of 2D MnSSe unit cell, and free atoms of Mn, S, and Se, respectively. The obtained $E_{cho}$=3.83 eV/atom very close to silicene (3.98 eV/atom) [33] and germane (3.26 eV/atom) [33]. Figure 1(b) displays the calculated phonon dispersion, in which no imaginary-vibration mode was found at the discretionary wave vector, indicating the dynamically stability of single-layer MnSSe. Thus, 2D MnSSe crystal with 1$T$ structure can be considered as stable.

For the fabrication of low-dimensional crystal-based electronics or spintronics devices, high in-plane stiffness is crucial to avoiding the curling or bulking of a 2D nanosheet. In-plane stiffness can be estimated by the 2D Young's modulus:

$$Y_{2D} = \frac{1}{S_0}\left(\frac{\partial^2 E_s}{\partial \varepsilon^2}\right)_{\varepsilon=0}, \quad (1)$$

$$\varepsilon=(l-l_0)/l_0, \quad (2)$$

where strain energy $E_s$ is the energy difference with reference to pristine MnSSe layer, $\varepsilon$ denotes the in-plane uniaxial strain. $l$ and $l_0$ denote the strained and unstrained lattice constants, respectively. $S_0$ is the unstrained surface area. Based on the optimized lattice constant 3.522 Å, we plotted Es versus ε in the range from -4% to 4% in Fig. 1(c). $Y_{2D}$ is estimated to be 53 N/m, which is very close to the predicated values of MnPSe$_3$ (36 N/m) [34], CrGeTe$_3$ (38.3 N/m) [35], Nb$_3$X$_8$ (X=Cl, Br, and I) (46~65 N/m) [36], and monolayer FeSe (80 N/m) [37]. Nevertheless, this value extracted here is significantly lower than that of graphene (~340 N/m) [38], indicating the soft and flexible nature of MnSSe monolayer that would facilitate its application under external strain.

Then, according to the elastic theory, one can derive the gravity-induced out-of-plane deformation h as follow [39]:

$$h/L \approx (\rho g L/Y_{2D})^{1/3}, \quad (3)$$

where $\rho$=2.57×10$^{-6}$ kg/m$^2$ is the density of the MnSSe monolayer, g is the gravitational acceleration (9.81 N/kg). For convenient comparison with the previous works [34-36], we take the size of the flake $L$=100 μm, resulting in $h/L$=3.63×10$^{-4}$, which is comparable to that of graphene [39]. Therefore, monolayer MnSSe hosts sufficient rigidity and can keeps free-standing planar structure without substrate.

**B. Intrinsic half-metallic ferromagnetism**

Spin-polarized calculations show that ferromagnetic (FM) state is 80 meV and 2.8 eV ones lower in energy than antiferromagnetic (AFM) and non-magnetic (NM) per unit cell, respectively. The net magnetic moment is 3 $\mu_B$ for a cell suggests a large spin polarization in this intrinsic ferromagnetic material. The spin polarizations are mainly dominated by Mn atoms [Fig. 2(a)], leading to a large magnetic moment of 3.654 $\mu_B$ per Mn atom, constants with the high spin state of $Mn^{3+}$. In contract, both S and Se atoms carry relatively small opposite moment (-0.223 $\mu_B$ for S and -0.322 $\mu_B$ for Se), the two chalcogenide atoms are antiferromagnetically coupled to the Mn atoms and hardly magnetized. The calculated electric band structure in displayed in Fig. 2(b) shows half-metallicity nature of MnSSe. Electrons of spin-down channel are insulating with a direct band gap of 1.14 eV at Γ point, while those of spin-up channel are metallic with a negative gap of -0.2 eV, leading to 100% spin-polarized conduction electrons. From the projected density of state, the bands of the spin-up channel near the Fermi level are predominantly contributed by the Mn, S, and Se states [Fig. 2 (c)].

The magnetocrystalline anisotropy energy (MAE) is the energy needs to overcome the "barrier" when moving the direction of a magnetic moment from the easy axis to the hard one, which determines the type of magnetic order at low temperature. Magnetic 2D materials with easy-plane magnetic anisotropy possess quasi-long-range ordered phase at low temperature. Meanwhile, Berezinskii-Kosterlitz-Thouless (BKT) transition occurs at crucial temperature with a change of an exponential law spin-spin correlation function behavior above $T_{BKT}$ to a power law one below it. By contrast, 2D system with an easy-axis magnetic anisotropy exhibit the so-called Curie transition or ferromagnetically transition to a long-range ordered low-temperature phase, *i.e.*, ferromagnetically ordered low-temperature phase [40]. MAE can be calculated as MAE=$E_{[uvw]}$-$E_{min}$, where $E_{min}$ denotes the energy of the most stable spin orientation. The polar angle dependence of MAE on the *xz*, *yz* and *xy* planes are illustrated in Figs. 3(a) and 3(b). we observed a strong dependence of the MAE on the $\theta$, while a negligible dependence on the $\varphi$. MAE is 0 at $\theta = 0°$ or $180°$, and reaches maximum value of 220 $\mu eV$ per Mn atom in *xy* plane, implying preferred out-of-plane axis. Therefore, 1*T*-MnSSe monolayer belongs to the family of 2D Ising magnets. Above zero temperature, the existence of heat fluctuation can destroy the long-range order, the large MAE found here will help to stabilize this ordering at a higher temperature.

To understand the temperature effect on the magnetism, we then performed the standard Monte Carlo Metropolis algorithm simulations based on the Ising Hamiltonian model,

$$\widehat{H} = -\sum_{i,j} J_{ij} M_i M_j, \tag{4}$$

where $J_{ij}$ represents the nearest-neighboring exchange coupling constant. $M_i$ and $M_j$ are the spin magnetic moments of nearest-neighbor unit cell. $J_0$ is defined as [41]

$$J = \frac{E_{ex}}{2z|S|^2}, \qquad (5)$$

where $E_{ex}$ is the exchange energy, the energy difference $\Delta E$ between AFM and FM configurations. $z = 6$ is the number of the nearest Mn neighbors. We obtain $J_0$=0.74 meV and predict a $T_c$ of about 72 K [(Fig. 3(c)]. This value is dramatically larger than the recent experimental reports of 2D CrI$_3$ (45 K) [42,43] and Cr$_2$Ge$_2$Te$_6$ (20 K) [44].

The microscopic origin of FM coupling in MnSSe can be understood by the competition of direct exchange interaction (Mn-Mn) and the super-exchange interaction (Mn-X-Mn, X= S or Se) mediated by the neighboring S and Se atoms, as shown in Fig. 3(d). For the direct interaction between the nearest neighbor Mn, their d orbitals overlap directly, which gives rise to AFM coupling. The Goodenough-Kanamori-Anderson (GKA) rule propose that FM coupling in systems with 90º bond-angles is energy favorable. According to Launay and Wagner the exchange integral $J$ in 2D materials has the approximate form of $J \approx 2k+4\beta S$ [45]. The first term $k$ is called the potential exchange, which is positive due to the Hund's first rule.[46] The second term consists of the hopping integral $\beta$ and overlap integral $S$. Since the bond-angles of Mn-S-Mn (94.8º) and Mn-Se-Mn (88.7º) are very close to 90º, the Mn-$d$ orbitals are therefore nearly orthogonal to the $p$ orbitals of S and Se, leading to a negligible overlap integral [Fig. 3(e)], and hence a positive $J \approx 2k$. As such, single layer MnSSe adopts a ferromagnetic ordering because of the relatively large FM super-exchange interaction.

### C. Carrier doping controlled magnetization transition

Recently, Jiang et al have experimentally demonstrated the controllability of the magnetic properties of both monolayer and bilayer CrI$_3$ by electrostatic doping, and observed electron carrier doping induced AFM-FM transition [47]. The modulation of the easy magnetization axis of Fe monolayer adsorbed on graphene substrate by charge injection was theoretically proposed [48]. A transition from ferromagnetic metal to half-metal was also predicated in ScCl exfoliated nanosheet with hole doping [49]. We considered carrier concentration up to 2.4×10$^{14}$ cm$^{-1}$ (0.1 electrons/holes per atom).

$\Delta E$ at various of carrier concentration is displayed in Fig. 4(a). Clearly, it substantial increases (decreases) linearly with doping, which mean that the exchange coupling can be efficiently enhanced or quenched with carrier doping. FM state is always the ground state over the whole doping range, implying that the interaction between moments has maintained FM state. This pronounced magnetic response toward carrier concentration could be attributed to the relatively enhanced through-space interactions and weakened through-bound interaction induced by carrier doping, since the crystal structure of MnSSe remains unchanged. The rapid and linear variation of magnetic response with carrier doping in MnSSe monolayer can be used to design magnetoelectric coupling spintronics devices.

On the other hand, carrier doping not only changes strength of magnetic coupling, but also affect MAE. Plotted in Fig. 4(b) is the total energy difference MAE=$E_{[001]}$-$E_{[100]}$ over a wide range of carrier concentrations. With electron charge doping, MnSSe retains its out-of-plane easy axis nature. In contrast, the sign of MAE is changed from positive to negative when a small amount of hole charge injected, implying possible change in easy magnetization situation. To characterize this change more visibly, as an example, we illustrated in Figs. 4(c) and 4(d) the MAEs of *xz*, *yz* and *xy* planes for hole-doped MnSSe. Obviously, MAE is rather strongly dependent on the out-of-plane angle $\theta$, and an easy *xy* plane is observed. Following the XY model [50], the critical temperature of the BKT transition can be calculated as

$$T_{BKT} = \frac{0.89J}{k_B}, \qquad (6)$$

where $k_B$ is the Boltzmann constant. *J* is the exchange parameter between adjacent spins, which can be evaluate from the $\Delta E$ as $J=\Delta E/8$ [51-53]. This estimation can be interpreted as following: there are 6 neighbors with the same spin in FM configuration, while 4 neighbors with the same spin and the other two with opposite direction in the AFM configuration, thus the magnetism energy for the two configurations can be expressed as $E_{FM}$=-6*J* and $E_{AFM}$=2*J*, respectively. As an example, at the hole doping concentration of $n$=2.23×10$^{14}$ cm$^{-2}$ (0.08 holes per atom), *J* is extracted to be 13.9 meV, which yields a $T_{BKT}$ of 145 K. This transition temperature is significantly higher than the liquid nitrogen temperature (77 K). For instance, BKT transition has already been observed or predicted in many other systems, including 2D magnetic crystal,[52-55] surface reconstructions [56], interface superconductors [57,58], and trapped atomic gas [59].

As discussed above, an unusual finding of electrically switchable transition between a low-temperature quasi-long-range ordered phase and a long-range ferromagnetic ordered one is constructed, in clear contrast to the relatively small changes of magnetic coupling strength and MAE predicated in other systems [42,60-62].

## D. Spin field effect transistor with giant magnetoresistance effect

The field-controllable spin-dependent transport is promising for spintronics. As to the applications, the doping-induced switch between the in-plane and out-of-plane directions of easy-axis can be used to design 2D magnetoelectric devices. As shown in Fig. 4(b), MnSSe is transformed from an out-of-plane ferromagnet to an in-plane one at hole doping with concentration above $n_c$=4.2×10$^{13}$ cm$^{-2}$, as suggested by the reversible easy-axis. In this case, the hetero-magnetic interface that composes these two states appears around this critical point. The strong interface scattering will lead to a sharp decrease in conductivity or an insulator transition. Below critical doping, in contrast, MnSSe hosts low resistance. Based on the above description and discussion, 2D MnSSe nanosheet was proposed to construct spin field effect transistor (spin-FET) with giant magnetoresistance effect. The device is schematically shown in Fig. 4(e). In short, MnSSe monolayer is double gated and directly sandwiched by top and bottom dielectric layers, *e.g.*, $SiO_2$/Si. The gate voltage $V_G$ is applied to control the carrier concentration and hence the easy axis. The source-drain voltage drives spin-polarized current.

One of the immediate consequences of injecting charge carrier in a crystal is to shifts the Fermi energy. We therefore track the evolution of the band structure in the whole range of doping. As expected, MnSSe shows a transition to a ferromagnetic metal with moderate hole doping $n=1.4\times10^{14}$ cm$^{-2}$ (0.5 holes per atom), see Fig. 5(a). Below this critical point, both carrier charge concentration and doping type have almost no effect on the size of the spin-gap [Fig. 5(b)], demonstrating the quite robust half-metallic ferromagnetism. Note that the critical doping for ferromagnetic metal is smaller than $n_c$, therefore, the proposed device could conduct a 100% spin-up current $I_{up}$. Due to the doping effect, the carrier charge concentration can be controlled towards or away from the $n_c$ by applied fields (voltages). The vertical electric field can act as a "gate" to switch the Iup [Fig. 4(f)], as such a spin-FET operated purely via field is constructed. Although a high-temperature or room temperature device and double spin-polarized currents in MnSSe monolayer have not been achieved here, it is meaningful to note that a rather strong magnetic coupling and bipolar control depending on crystal structure, constituent elements, and external interaction, *etc*. [63-66]. This leaves room for material selections. Previously, a modulation with charge density injected by using ionic liquid as gate dielectric at the level of ~$10^{15}$ cm$^{-2}$ has been reported in many 2D materials [67-69]. So, it is experimentally feasible to tune the magnetism by carrier doping and realize the 2D spin-FET proposed here.

**E. Unique intralayer charge transfer**

Since carrier doping is mainly implemented by the liquid (or solid ionic) gating technique in experiments [70,71], where an external electric field is applied generally across the nanosheet. Therefore, it is necessary to find out the effect of this field on the physical properties of the material. As shown in Fig. 6(a), the field is applied parallel to the z-axis, pointing from S- to Se-layer as indicated by the red arrow. The total energy of AFM and FM configurations with respect to $E_\perp$ are plotted in Fig. 6(b). The energy of the NM configuration is much higher than those of the magnetic ones, so it is not shown here. Both EAFM and EFM decrease in a quadratic way while the field increases away from zero, forming a dome-like shape, implying that $E_\perp$ makes them more stable, in sharp contrast to the monotonically increased tendency theoretical reported in 2D nanosheet under strain [72,73]. The FM state always has lower energy than the AFM one, suggesting that the FM ordering is robust within a wide range of fields. Our test calculations also reveal that the crystal structure, $\Delta E$, and band structure are minimally changed under a rather strong field. For instance, the influence of fields on crystal structure and energy has proved to be negligible [20,74].

Finally, we focus on the intra-layer charge transfer behavior of Janus MnSSe monolayer. The Bader charge analysis shows that both the electron number of S and Se [S ($N_S$) and Se ($N_{Se}$)] exhibit an approximately linear relation with external field in the range from $E_\perp$=-3 to 3 V/Å, while that of Mn ($N_{Mn}$) keeps mostly unchangeable [Fig. 6(c)]. This unique intralayer transfer constitutes the second major finding in this study that has not been observed in TMDs previously. Moreover, $N_S$ and $N_{Se}$ varied reversely, implying that the charge transfer mainly occurs between S and Se atoms. What is far more important is that the electrons under external electric field are usually said to flow from the negative to the positive side of the field, which is opposite to the observation above, as when the field applied from S to Se layer, *i.e.*, the positive electric field we defined in Fig. 6(a), $N_S$ decreases with the increasing of $E_\perp$, contrarily, $N_{Se}$ increases with increasing $E_\perp$. To gain deeper insights into this anomalous behavior, we examine the bonding character by the electron localization function (ELF), which indicates the degree of electron localization. For clearly, the (110) plane catting through two S atoms, two Se atoms, and two Mn atoms was selected and displayed in Fig. 6(d). Electrons are mainly localized around S and Se sites but rarely between S/Se and Mn atoms, suggesting the ionic character of Mn-S and Mn-Se bonds.

Due to the mirror asymmetry or electronegativity difference between S and Se atoms, $N_{Se}$ is 0.2 |e| less than $N_S$ when $E_\perp$=0 V/Å, which may result in a built-in electric field, $E_{in}$. In support of this conjecture, we plot in Fig. 6(e) the planar averaged electrostatic potential energy along the z-direction for MnSSe without external field. An intralayer potential gradient with a height of 2.66 eV is observed, indicating the existence of a net electric field pointing from Se- to S-layer. This indicates that this anomalous charge transfer is a joint effort of the external and in the built-in electric fields. The increased $E_\perp$, with the positive field as a representative, weeks $E_{in}$ and hence pushes electrons from S to Se. In addition, $N_S$ departs from the linear relationship from $E_\perp$=-1 V/Å and manifests as a dip cantered at -1.3 V/Å. This echo with a dip at the same field only appears in NS but not in $N_{Se}$ that needs further study.

## IV. CONCLUSIONS
In summary, we have reported a promising alternative for tunning the magnetic coupling and electric structure in the monolayer MnSSe. The intrinsic ferromagnetic half-metallicity with a Curie temperature around 72 K is predicated. We demonstrated that carrier doping can sensitively enhance or quench this ferromagnetic coupling. The magnetization easy axis can be easily tuned between out-of-plane and in-plane with a small amount of hole doping. These findings render MnSSe monolayer great potential for application in electrically controllable spintronic devices, such as spin-FET with giant magnetoresistance effect. We also found a pronounced intralayer charge density response towards external electric field for the first time, which is intriguing and could be considered as a new degree of freedom that will promotes the development of nanoscale electronics and spintronics devices on the basis of 2D materials with asymmetry structure.

**Acknowledgements**

Part of this research used resources of National Supercomputer Center in LvLiang of China. G.Y. thanks H. M. Zhang for fruitful discussions. This work was supported by the Scientific Research Project of Inner Mongolia University of Technology (Grant No. BS2021058) and National Natural Science Foundation of China (Grant No. 12104294).

**References**
[1] A. Fert, Rev. Mod. Phys. **80**, 1517-1530 (2008).
[2] S. A. Wolf, D. D. Awschalom, R. A. Buhrman, J. M. Daughton, S. von Molnár, M. L. Roukes, A. Y. Chtchelkanova, and D. M. Treger, Science **294**, 1488 (2001).
[3] Q. H. Wang, K. Kalantar-Zadeh, A. Kis, J. N. Coleman, and M. S. Strano, Nat. Nanotech. **7**, 699-712 (2012).
[4] M. Gibertini, M. Koperski, A. F. Morpurgo, and K. S. Novoselov, Nat. Nanotech. **14**, 408-419 (2019).
[5] H. Yuan, M. S. Bahramy, K. Morimoto, S. Wu, K. Nomura, B.-J. Yang, H. Shimotani, R. Suzuki, M. Toh, C. Kloc *et al.*, Nat. Phys. **9**, 563-569 (2013).
[6] S. Wu, J. S. Ross, G.-B. Liu, G. Aivazian, A. Jones, Z. Fei, W. Zhu, D. Xiao, W. Yao, D. Cobden *et al.*, Nat. Phys. **9**, 149-153 (2013).
[7] Y. C. Cheng, Z. Y. Zhu, M. Tahir, and U. Schwingenschlögl, EPL (Europhysics Letters) **102**, 57001 (2013).
[8] A.-Y. Lu, H. Zhu, J. Xiao, C.-P. Chuu, Y. Han, M.-H. Chiu, C.-C. Cheng, C.-W. Yang, K.-H. Wei, Y. Yang *et al.*, Nat. Nanotech. **12**, 744-749 (2017).
[9] J. Zhang, S. Jia, I. Kholmanov, L. Dong, D. Er, W. Chen, H. Guo, Z. Jin, V. B. Shenoy, L. Shi *et al.*, ACS Nano **11**, 8192-8198 (2017).
[10] X. Wang, D. Chen, Z. Yang, X. Zhang, C. Wang, J. Chen, X. Zhang, and M. Xue, Adv. Mater. **28**, 8645-8650 (2016).
[11] M. L. Wenjun Liu, Ximei Liu, Ming Lei, Zhiyi Wei, Opt. Lett. **45**, 419-422 (2020).
[12] X. H. Wenzhou Chen, Xingqiang Shi, Hui Pan, ACS Appl. Mater. Interfaces **10**, 35289-35295 (2018).

[13] Y. Guo, S. Zhou, Y. Bai, and J. Zhao, Appl. Phys. Lett. **110**, 163102 (2017).
[14] J. He, and S. Li, Comput. Mater. Sci. **152**, 151-157 (2018).
[15] J. Wang, B. Lian, and S.-C. Zhang, Phys. Rev. Lett. **115**, 036805 (2015).
[16] T. M. R. Wolf, J. L. Lado, G. Blatter, and O. Zilberberg, Phys. Rev. Lett. **123**, 096802 (2019).
[17] H. Wang, J. Qi, and X. Qian, Appl. Phys. Lett. **117**, 083102 (2020).
[18] S. Jiang, J. Shan, and K. F. Mak, Nat. Mater. **17**, 406-410 (2018).
[19] Y. Y. Sun, L. Q. Zhu, Z. Li, W. Ju, S. J. Gong, J. Q. Wang, and J. H. Chu, J. Phys. Condens. Matter. **31**, 205501 (2019).
[20] S. J. Gong, C. Gong, Y. Y. Sun, W. Y. Tong, C. G. Duan, J. H. Chu, and X. Zhang, Proc. Natl. Acad. Sci. U.S.A. **115**, 8511-8516 (2018).
[21] M. Weisheit, S. Fähler, A. Marty, Y. Souche, C. Poinsignon, and D. Givord, Science **315**, 349-351 (2007).
[22] J. Yuan, Y. Yang, Y. Cai, Y. Wu, Y. Chen, X. Yan, and L. Shen, Phys. Rev. B **101**, 094420 (2020).
[23] P. E. Blöchl, Phys. Rev. B **50**, 17953-17979 (1994).
[24] G. Kresse, and D. Joubert, Phys. Rev. B **59**, 1758-1775 (1999).
[25] G. Kresse, and J. Furthmüller, Phys. Rev. B **54**, 11169-11186 (1996).
[26] J. P. Perdew, K. Burke, and M. Ernzerhof, Phys. Rev. Lett. **77**, 3865-3868 (1996).
[27] S. L. Dudarev, G. A. Botton, S. Y. Savrasov, C. J. Humphreys, and A. P. Sutton, Phys. Rev. B **57**, 1505-1509 (1998).
[28] J. He, P. Lyu, L. Z. Sun, Á. Morales García, and P. Nachtigall, J. Mater. Chem. C **4**, 6500-6509 (2016).
[29] G. J. Ackland, M. C. Warren, and S. J. Clark, J. Phys. Condens. Matter. **9**, 7861-7872 (1997).
[30] A. Togo, and I. Tanaka, Scripta Materialia **108**, 1-5 (2015).
[31] K. Momma, and F. Izumi, J. Appl. Crystallogr. **44**, 1272-1276 (2011).
[32] V. Wang, N. Xu, J.-C. Liu, G. Tang, and W.-T. Geng, Comp. Phys. Commun. **267**, 108033 (2021).
[33] L. M. Yang, I. A. Popov, T. Frauenheim, A. I. Boldyrev, T. Heine, V. Bacic, and E. Ganz, Phys. Chem. Chem. Phys. **17**, 26043-26048 (2015).
[34] X. Li, X. Wu, and J. Yang, J. Am. Chem. Soc. **136**, 11065-11069 (2014).
[35] X. Li, and J. Yang, J. Mater. Chem. C **2**, 7071 (2014).
[36] J. Jiang, Q. Liang, R. Meng, Q. Yang, C. Tan, X. Sun, and X. Chen, Nanoscale **9**, 2992-3001 (2017).
[37] F. Zheng, J. Zhao, Z. Liu, M. Li, M. Zhou, S. Zhang, and P. Zhang, Nanoscale **10**, 14298-14303 (2018).
[38] A. Politano, and G. Chiarello, Nano Res. **8**, 1847-1856 (2015).
[39] T. J. Booth, P. Blake, R. R. Nair, D. Jiang, E. W. Hill, U. Bangert, A. Bleloch, M. Gass, K. S. Novoselov, M. I. Katsnelson *et al.*, Nano Lett. **8**, 2442-2446 (2008).
[40] J. M. Kosterlitz, and D. J. Thouless, J. Phys. C: Solid State Phys. **6**, 1181-1203 (1973).
[41] V. V. Kulish, and W. Huang, J. Mater. Chem. C **5**, 8734-8741 (2017).
[42] L. Webster, and J.-A. Yan, Phys. Rev. B **98**, 144411 (2018).

[43] B. Huang, G. Clark, E. Navarro-Moratalla, D. R. Klein, R. Cheng, K. L. Seyler, D. Zhong, E. Schmidgall, M. A. McGuire, D. H. Cobden *et al.*, Nature (London) **546**, 270-273 (2017).

[44] C. Gong, L. Li, Z. Li, H. Ji, A. Stern, Y. Xia, T. Cao, W. Bao, C. Wang, Y. Wang *et al.*, Nature (London) **546**, 265-269 (2017).

[45] J.-P. Launay, and M. Verdaguer. *Electrons in Molecules: From Basic Principles to Molecular Electronics*. 1st edn,  (Oxford University Press, 2013).

[46] P. A. Cox. *Transition Metal Oxides. An Introduction to their Electronic Structure and Properties*.  (Oxford University Press, 2010).

[47] S. Jiang, L. Li, Z. Wang, K. F. Mak, and J. Shan, Nat. Nanotech. **13**, 549-553 (2018).

[48] S. J. Gong, C.-G. Duan, Z.-Q. Zhu, and J.-H. Chu, Appl. Phys. Lett. **100**, 122410 (2012).

[49] B. Wang, Q. Wu, Y. Zhang, Y. Guo, X. Zhang, Q. Zhou, S. Dong, and J. Wang, Nanoscale Horiz **3**, 551-555 (2018).

[50] J. F. Fernández, M. F. Ferreira, and J. Stankiewicz, Phys. Rev. B **34**, 292-300 (1986).

[51] H. L. Zhuang, and R. G. Hennig, Phys. Rev. B **93**, 054429 (2016).

[52] D. Dey, and A. S. Botana, Phys. Rev. Mater. **4**, 074002 (2020).

[53] M. Ashton, D. Gluhovic, S. B. Sinnott, J. Guo, D. A. Stewart, and R. G. Hennig, Nano Lett. **17**, 5251-5257 (2017).

[54] A. N. Ma, P. J. Wang, and C. W. Zhang, Nanoscale **12**, 5464-5470 (2020).

[55] A. Bedoya-Pinto, J.-R. Ji, A. Pandeya, P. Gargiani, M. Valvidares, P. Sessi, F. Radu, K. Chang, and S. Parkin, arXiv:2006.07605 (2020).

[56] D. H. Baek, J. W. Chung, and W. K. Han, Phys. Rev. B **47**, 8461-8464 (1993).

[57] W. H. Zhang, Y. Sun, J. S. Zhang, F. S. Li, M. H. Guo, Y. F. Zhao, H. M. Zhang, J. P. Peng, Y. Xing, H. C. Wang *et al.*, Chin. Phys. Lett. **31**, 017401 (2014).

[58] N. Reyren, S. Thiel, A. D. Caviglia, L. F. Kourkoutis, G. Hammerl, C. Richter, C. W. Schneider, T. Kopp, A. S. Ruetschi, D. Jaccard *et al.*, Science **317**, 1196-1199 (2007).

[59] Z. Hadzibabic, P. Krüger, M. Cheneau, B. Battelier, and J. Dalibard, Nature (London) **441**, 1118-1121 (2006).

[60] Z. Guan, and S. Ni, ACS Appl. Mater. Interfaces **12**, 53067-53075 (2020).

[61] Q. Sun, S. Kwon, M. Stamenova, S. Sanvito, and N. Kioussis, Phys. Rev. B **101**, 134419 (2020).

[62] M. Abdollahi, and M. Bagheri Tagani, J. Mater. Chem. C **8**, 13286-13296 (2020).

[63] X. Li, X. Wu, Z. Li, J. Yang, and J. G. Hou, Nanoscale **4**, 5680-5685 (2012).

[64] R. Chua, J. Zhou, X. Yu, W. Yu, J. Gou, R. Zhu, L. Zhang, M. Liu, M. B. H. Breese, W. Chen *et al.*, Adv Mater, e2103360 (2021).

[65] S. Fu, K. Kang, K. Shayan, A. Yoshimura, S. Dadras, X. Wang, L. Zhang, S. Chen, N. Liu, A. Jindal *et al.*, Nat Commun **11**, 2034 (2020).

[66] C. Tang, K. K. Ostrikov, S. Sanvito, and A. Du, Nanoscale Horiz **6**, 43-48 (2021).

[67] H. Yuan, H. Shimotani, A. Tsukazaki, A. Ohtomo, M. Kawasaki, and Y. Iwasa, Adv. Funct. Mater. **19**, 1046-1053 (2009).

[68] A. S. Dhoot, C. Israel, X. Moya, N. D. Mathur, and R. H. Friend, Phys. Rev. Lett. **102**, 136402 (2009).

[69] J. T. Ye, S. Inoue, K. Kobayashi, Y. Kasahara, H. T. Yuan, H. Shimotani, and Y. Iwasa, Nat. Mater. **9**, 125-128 (2010).

[70] B. Lei, J. H. Cui, Z. J. Xiang, C. Shang, N. Z. Wang, G. J. Ye, X. G. Luo, T. Wu, Z. Sun, and X. H. Chen, Phys. Rev. Lett. **116**, 077002 (2016).

[71] T. P. Ying, M. X. Wang, X. X. Wu, Z. Y. Zhao, Z. Z. Zhang, B. Q. Song, Y. C. Li, B. Lei, Q. Li, Y. Yu *et al.*, Phys. Rev. Lett. **121**, 207003 (2018).

[72] Y. Ma, Y. Dai, M. Guo, C. Niu, Y. Zhu, and B. Huang, ACS Nano **6**, 1695-1701 (2012).

[73] Y. Zhou, Z. Wang, P. Yang, X. Zu, L. Yang, X. Sun, and F. Gao, ACS Nano **6**, 9727-9736 (2012).

[74] Q.-F. Yao, J. Cai, W.-Y. Tong, S.-J. Gong, J.-Q. Wang, X. Wan, C.-G. Duan, and J. H. Chu, Phys. Rev. B **95**, 165401 (2017).

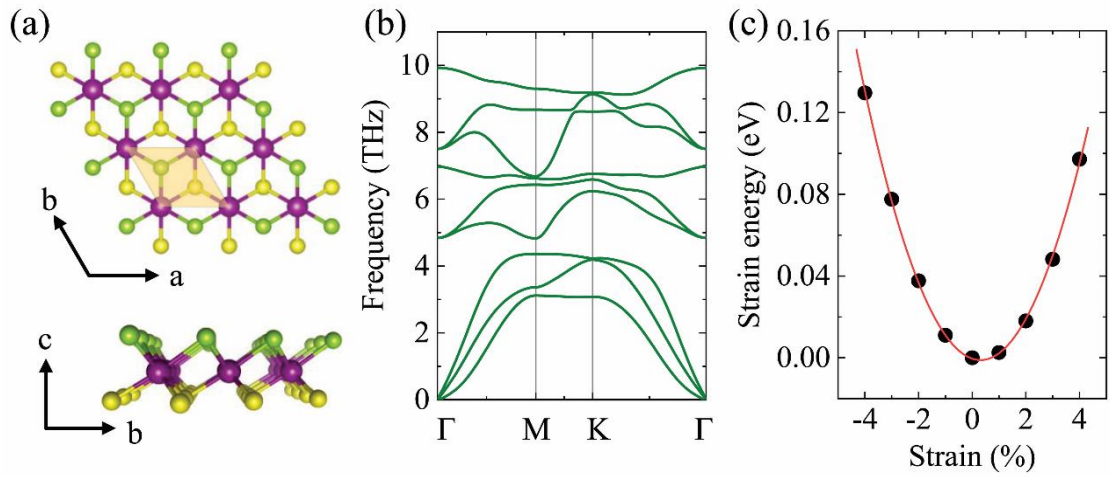

FIG. 1. (a) Top and side view of a MnSSe monolayer with 1T phase, where Mn, S, and Se atoms are purple, yellow, and green, respectively. The orange shaded region indicates the primitive unit cell. (b) Calculated phonon dispersion for 2D MnSSe crystal. (c) The strain energy of MnSSe monolayer under uniaxial strain.

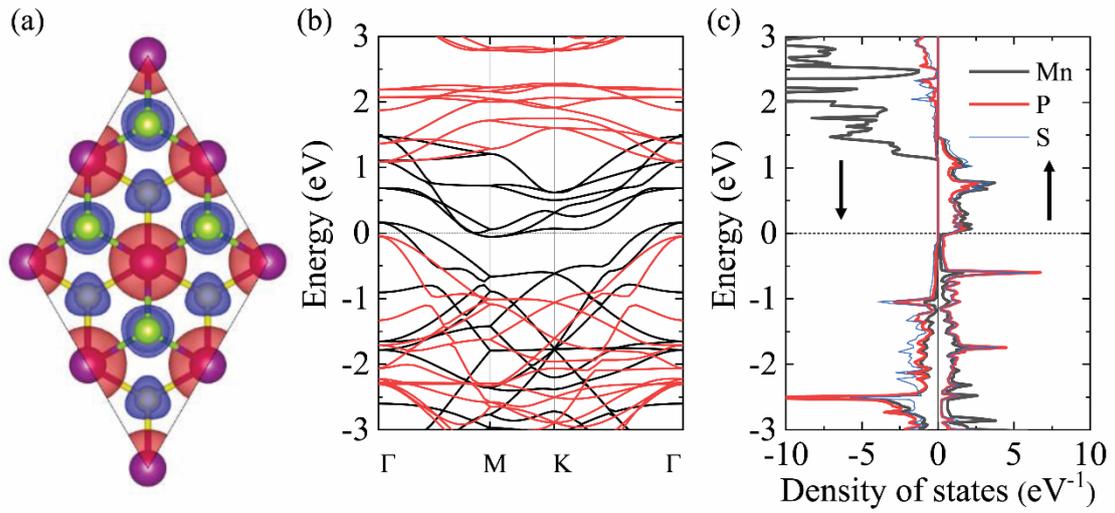

FIG. 2. (a) Spin-resolved charge density with an iso-value of 0.01 e/Å3 in the FM state. Red (blue) ones indicate the positive (negative) values. (b) Spin-resolved electronic band structures. The black (red) line represents the spin-up (spin-down) bands. (c) Projected density of state. In (b) and (c), the Fermi level is denoted by the dotted line at zero.

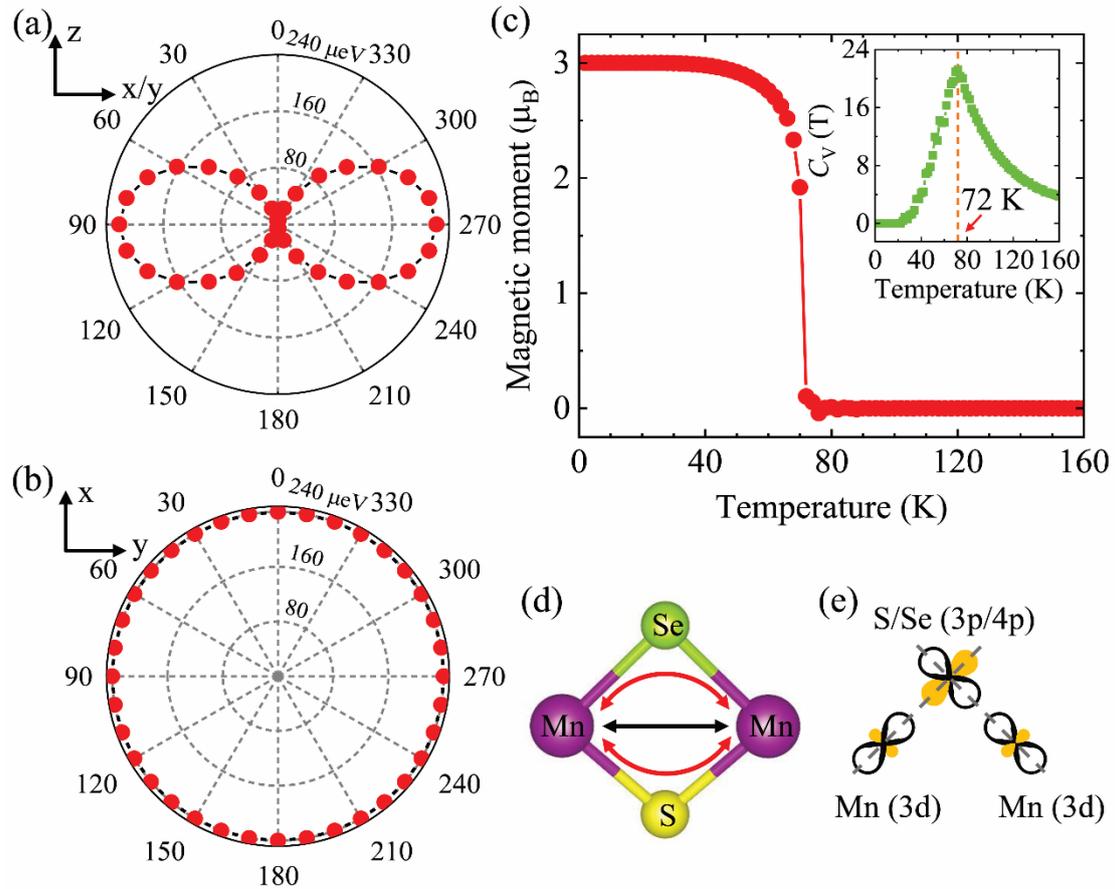

FIG. 3. (a, b) Dependence of the MAE per Mn atom for pristine single-layer MnSSe on the out-of-plane polar angle $\theta$ and in-plane azimuth angle $\varphi$, respectively. (c) The simulated magnetic moment and specific heat ($C_v$) with respect to temperature. (d) Schematic mechanism of direct Mn-Mn (black arrows) and super-exchange Mn-S(Se)-Mn (red arrows) interactions. (e) Mechanism of super-exchange interaction for a nearly 90° Mn-S(Se)-Mn bond angle.

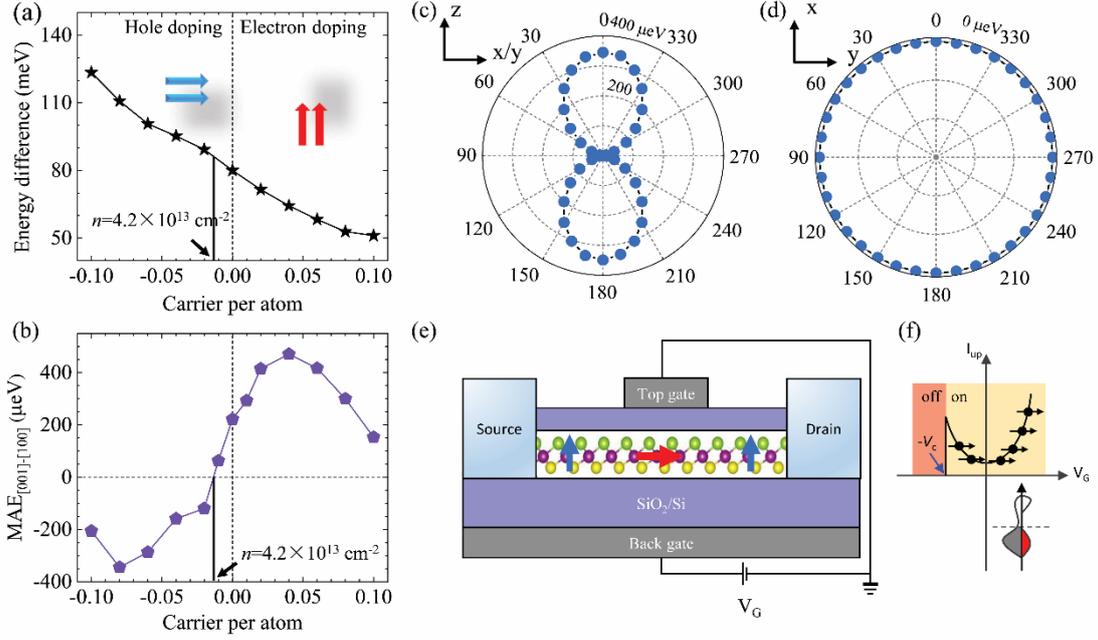

FIG. 4. (a) ΔE (=$E_{AFM}$-$E_{FM}$, in meV per unit cell) and (b) MAE=$E_{[001]}$-$E_{[100]}$ plotted as a function of carrier concentration up to $2.8\times10^{14}$ cm$^{-2}$ (0.1 electrons per atom). (c) and (d) Dependence of the MAE on the out-of-plane polar angle $\theta$ and in-plane azimuth angle $\varphi$ for MnSSe with carrier concentration of $n$=0.08 holes per atom. (e) A schematic illustration of the proposed spin-FET with giant magnetoresistance effect based on the 2D MnSSe crystal. The gate voltage $V_G$ is applied to control the carrier concentration and easy axis. (f) Schematic plot of the spin-down current $I_{up}$ versus $V_G$, with -$V_c$ indicates the voltage that can introduce a concentration of $n_c$.

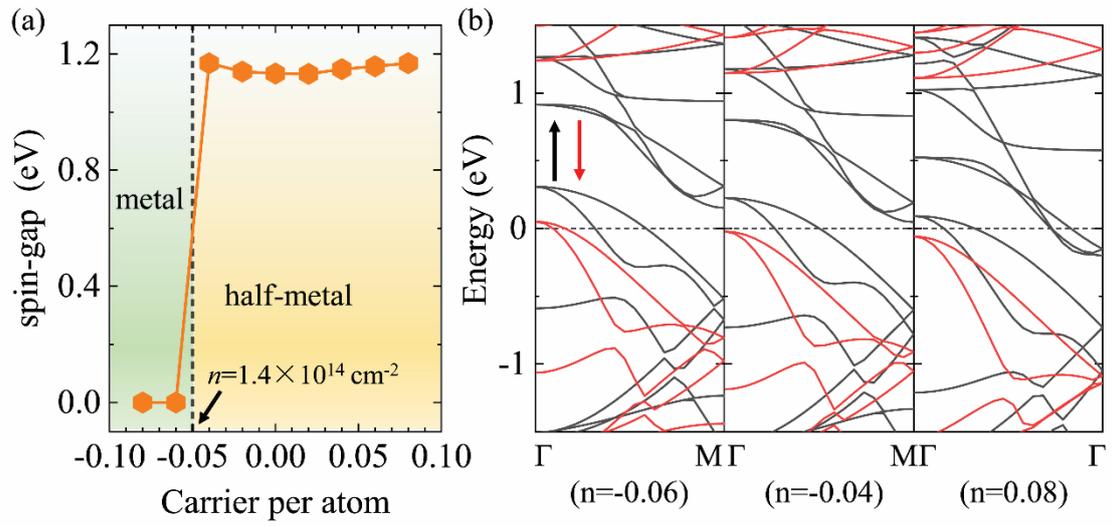

FIG. 5. (a) Variation of the band gap of spin-down channel with charge carrier concentration. (b) Three typical spin-polarized band structures.

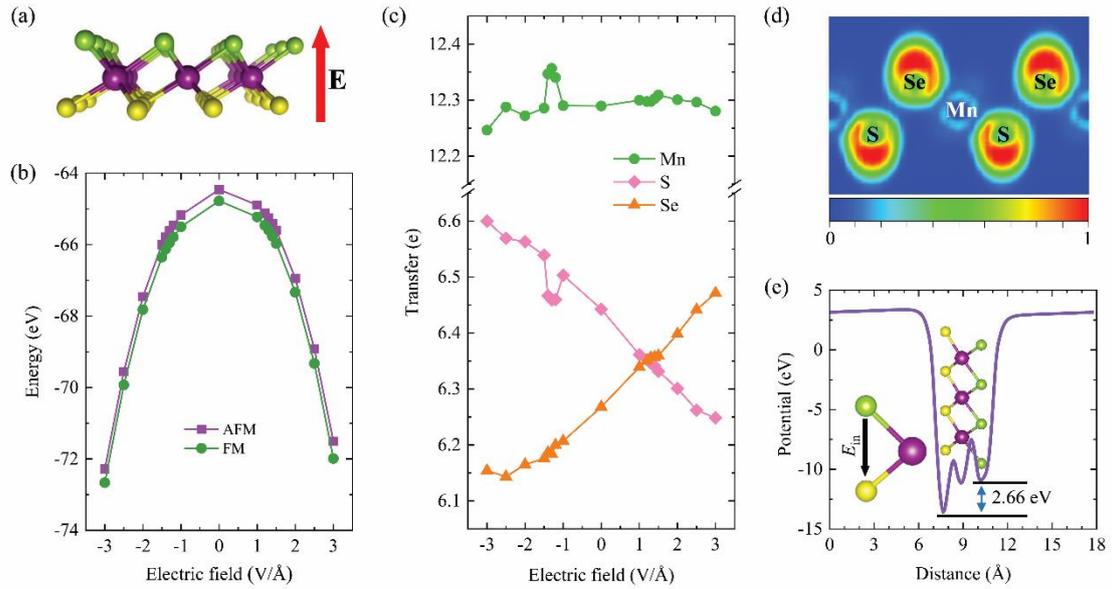

FIG. 6. (a) Scheme of the relation between MnSSe nanosheet and the applied external electric field $E_\perp$. The red arrow indicates the direction of both the positive electric field and the $z$-axis. (c) System energies for AFM and FM configurations versus $E_\perp$, both with a dome shape. (c) Variation of the total charge of Mn, S, and Se atoms with respect to $E_\perp$. (d) ELF in (110) plane under zero filed. (e) Planar average of the electrostatic potential energy for MnSSe along the $z$-direction under zero filed. The inset shows the scheme of built-in electric field $E_{in}$.